\documentclass[letterpaper, 10 pt, conference]{ieeeconf}  

\IEEEoverridecommandlockouts                              

\usepackage{breakcites}
\usepackage{gensymb}
\usepackage{amsmath} 
\usepackage{amssymb}  
\usepackage[dvipsnames]{xcolor}
\usepackage{makecell}
\usepackage{booktabs}
\usepackage{graphicx}
\usepackage{comment}
\usepackage{subfigure}
\usepackage{cite}
\usepackage{url}

\makeatletter
\let\NAT@parse\undefined
\makeatother
\usepackage{hyperref}
\usepackage{cleveref}
\usepackage{fancyhdr}
\newcommand{\norm}[1]{\left\lVert#1\right\rVert}
\newcommand{\abs}[1]{\left|#1\right|}

\fancypagestyle{firstpage}{\lhead{\small \textbf{{\fontfamily{cmss}\selectfont
2024 American Control Conference (ACC)\\July 8-12, 2024, Toronto, Canada}}}}

\title{\Large \bf
Model Predictive Control of Diesel Engine Emissions Based on Neural Network Modeling
}

\author{Jiadi Zhang$^{1}$, Xiao Li$^{1}$, Ilya Kolmanovsky$^{1}$, Munechika Tsutsumi$^{2}$, and Hayato Nakada$^{2}$ 
\thanks{$^{1}$J. Zhang, Xiao and I. Kolmanovsky are with the Department of Aerospace Engineering, University of Michigan, Ann Arbor, MI 48109, USA
        {\tt\small \{jiadi,hsiaoli,ilya\}@umich.edu}}%
\thanks{$^{2}$H. Nakada and M. Tsutsumi are with Hino Motors, Ltd., Tokyo 191-8660, Japan
       {\tt\small \{mu.tsutsumi,hayato.nakada\}@hino.co.jp}}%
}
\renewcommand{\baselinestretch}{0.85}
\begin{document}

\maketitle
\thispagestyle{firstpage}


\begin{abstract}
This paper addresses the control of diesel engine nitrogen oxides (NOx) and Soot emissions through the application of Model Predictive Control (MPC). 
The developments described in the paper are based on a high-fidelity model of the engine airpath and torque response in GT-Power, which is extended with a feedforward neural network (FNN)-based model of engine out (feedgas) emissions identified from experimental engine data to enable the controller co-simulation and performance verification. A Recurrent Neural Network (RNN) is then identified for use as a prediction model in the implementation of a nonlinear economic MPC that adjusts intake manifold pressure and EGR rate set-points to the inner loop airpath controller as well as the engine fueling rate. Based on GT-Power engine model and FNN emissions model, the closed-loop simulations of the control system and the plant model, over different driving cycles, demonstrate the capability to shape engine out emissions response by adjusting weights and constraints in economic MPC formulation. 
\end{abstract}

\section{Introduction} \label{sec:intro}

Important priorities in diesel engine control development include the reduction of engine emissions and establishing a streamlined and systematic controller implementation and calibration process \cite{ashok2016review,dewangan2020combustion}. Further opportunities to address these priorities emerge due to rapid advances in machine learning, artificial intelligence, and optimization-based control methods while exploiting growing onboard computing power\cite{hafner2000fast,liao2020model}.

This paper focuses on the interplay between neural networks-based modeling and Model Predictive Control (MPC).  A high-fidelity model of the engine airpath and torque response is first augmented with a feedforward neural network (FNN)-based model of engine out (feedgas) emissions.  The FNN is identified from experimental engine data, enabling controller co-simulation and performance verification. While the FNN approach to modeling engine emissions has also been adopted in our previous work on LPV MPC \cite{IFAC2023}, in this paper a systematic methodology is proposed and used to tune hyperparameters of the FNN (such as the learning rate and momentum constants in the stochastic gradient descent) to enhance the emissions prediction accuracy. The process we used in informing a single training data set for FNN by merging data sets resulting from open-loop steady-state dynamometer experiments and closed-loop transient experiments (with an existing baseline controller) has also been modified and we report it in detail.

A Recurrent Neural Network (RNN) is then identified for use as an MPC prediction model.  This RNN has a small number of inputs representing signals available for engine control and is simpler as compared to FNN, thereby facilitating MPC computational implementation.  We adopt a ``modular'' identification process that exploits an abstraction of the airpath controller to uncouple the identification of the RNN model for emissions from the need to have a fully calibrated airpath controller. 

\begin{figure}[ht!]
\centering
\includegraphics[width=0.45\textwidth]{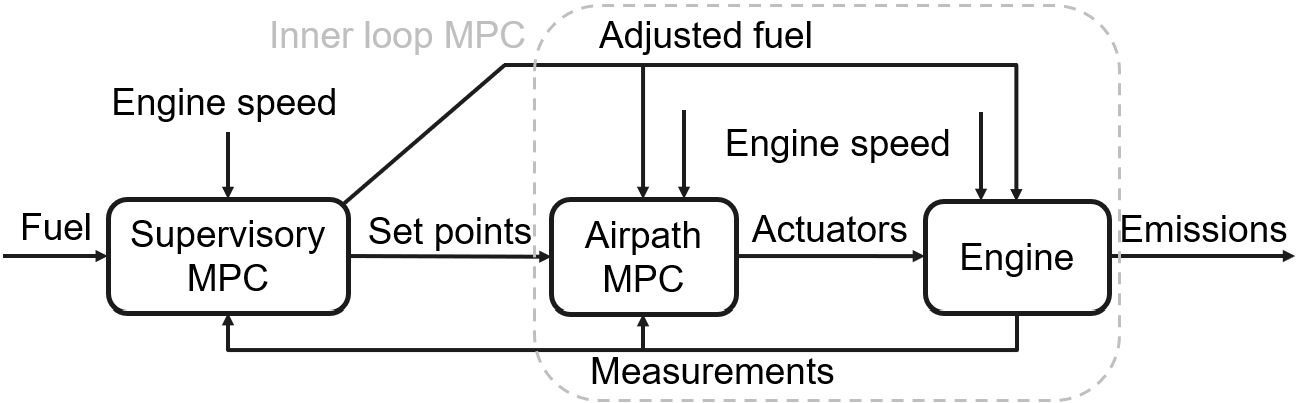}
\caption{A high-level control architecture diagram. }
\label{fig:high-level structure}
\end{figure}

The economic MPC (EMPC) is designed next based on the RNN to adjust intake pressure and EGR rate set-points (targets) to the inner loop airpath controller as well as the fueling rate. The overall control system  (see Figure~\ref{fig:high-level structure}) integrates this outer-loop EMPC with an inner-loop MPC for the airpath system that we refer to as the airpath MPC. 
Such a hierarchical MPC architecture, which is the focus in this paper, is adopted following 
\cite{liao2020model},
where its capability to improve transient engine control and reduce emissions has been demonstrated.
However, our implementation of EMPC is novel as it exploits RNN as the prediction model, and a different formulation of the cost and constraints which explicitly reflect predicted $NOx$ and $Soot$ feedgas emissions.
The closed-loop simulations of the designed controller and the plant model based on GT-Power engine model and FNN emissions model over FTP and WHTC drive cycles that will be reported demonstrate the capability to shape engine out emissions response by adjusting weights and constraints in EMPC formulation.

 The airpath MPC is used to follow the intake manifold pressure and EGR rate set-points computed by EMPC.
 The airpath MPC
 adopted in this paper 
 exploits a combination of feedforward MPC and feedback rate-based MPC following \cite{liao2020model,ZHANG2022181} to control Exhaust Gas Recirculation (EGR) valve and Variable Geometry Turbocharger (VGT) vane actuators of the engine (see Figure~\ref{fig:engine}).
 In automotive applications, a feedback tracking controller is typically integrated with a feedforward controller which relies on look-up tables for set-points and actuator positions
 to enhance transient response (\cite{norouzi2021model}).
 The utilization of an MPC-based feedforward strategy  \cite{liao2020model, ZHANG2022181}, as an alternative to a look-up table based implementation,  has demonstrated further improvements in closed-loop transient response.
 
The development of a different neural network-based MPC solution for diesel engines is pursued in \cite{gordon2022end}, where the focus is on control of engine fueling to reduce emissions.  Our paper is distinguished by its focus on a hierarchical MPC architecture, where EMPC is used as a supervisor adjusting set-points to the airpath MPC, concurrently with adjusting/limiting main injection fuel quantity. We also exploit a relatively simple RNN architecture as the prediction model in EMPC versus LSTM in \cite{gordon2022end}.

The paper is organized as follows.  The FNN and RNN 
emission modeling is addressed in Section~\ref{sec:modeling}. The EMPC design is described in Section~\ref{sec:controller design}.  Section~\ref{sec:simulations} reports simulation results.  Finally, Section~\ref{sec:conclusion} presents concluding remarks.

\section{Diesel Engine Modeling}\label{sec:modeling} 

\begin{figure}[ht!]
\centering
\includegraphics[width=0.45\textwidth]{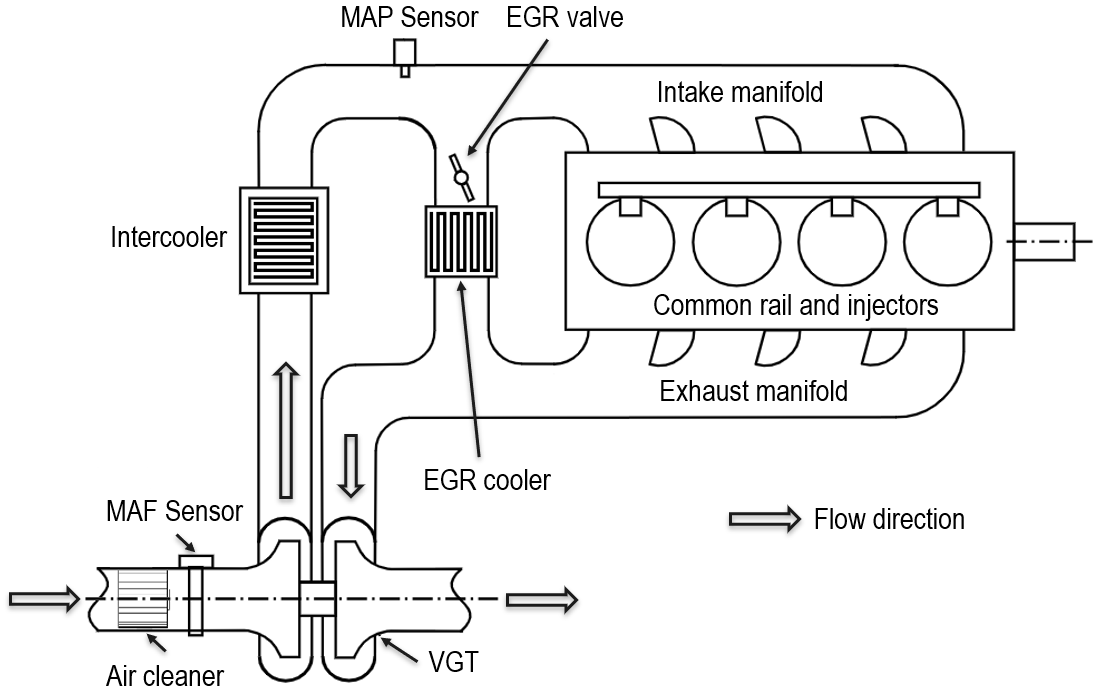}
\caption{Diesel engine with exhaust gas recirculation (EGR) and a variable geometry turbocharger (VGT).}
\label{fig:engine}
\end{figure}

The diesel engine schematics are shown in Figure~\ref{fig:engine} which illustrates the primary engine components, including a cylinder block, intake manifold, exhaust manifold, an Exhaust Gas Recirculation (EGR) system, and a variable geometry turbocharger (VGT) consisting of a turbine with the adjustable guide vanes and a compressor on the same shaft. 
The EGR system is used to reduce oxides of nitrogen ($NOx$) emissions.
It achieves this by actuating EGR valve that recirculates the exhaust gas into the intake manifold ($w_{\tt egr}$).  Once in-cylinder, recirculated burnt gas acts as an inert gas, and hence reduces peak in-cylinder temperatures and  $NOx$ formation. However, excessive EGR can lead to an increase in $Soot$ emissions and engine roughness.  There are strong dynamic interactions between EGR valve and VGT actuators in their effect on engine variables. In particular, the flows through EGR valve and VGT depend on the exhaust pressure which in turn is influenced by these flows. Similarly, both of these actuators affect the intake manifold pressure and the air-to-fuel ratio.

The airpath controller regulates the intake manifold pressure
($p_{\tt im}$) and EGR rate
($\chi_{\tt egr}$) to set-points (target values), where  the
EGR rate is defined by
\begin{equation} \label{eq:EGR_rate_def}
\chi_{\tt egr} = \frac{w_{\tt egr}}{w_{\tt egr} + w_{\tt c}},
\end{equation}
and where $w_{\tt egr}$ is the flow through the EGR valve and 
$w_{\tt c}$ is the flow through the compressor.
The intake manifold pressure and EGR rate set-points, $p_{\tt im}^{\tt adj}$ and $\chi_{\tt egr}^{\tt adj}$, respectively, are computed and provided to the airpath controller by the EMPC controller.

The starting point for the developments in this paper is a  high-fidelity diesel engine airpath model in GT-Power.  This model represents the responses of flows and pressures in different parts of the engine at a crankangle resolution, and it has been validated against experimental data from engine dynamometer testing.   This model has been extended in our work by an engine feedgas emissions model.  For the latter, we used available data sets from engine dynamometer testing and implemented an FNN.

\subsection{Data-driven Engine Feedgas Emissions Modeling}\label{subsec:fnn_modeling}

\begin{figure}[ht!]
\centering
\includegraphics[width=0.45\textwidth]{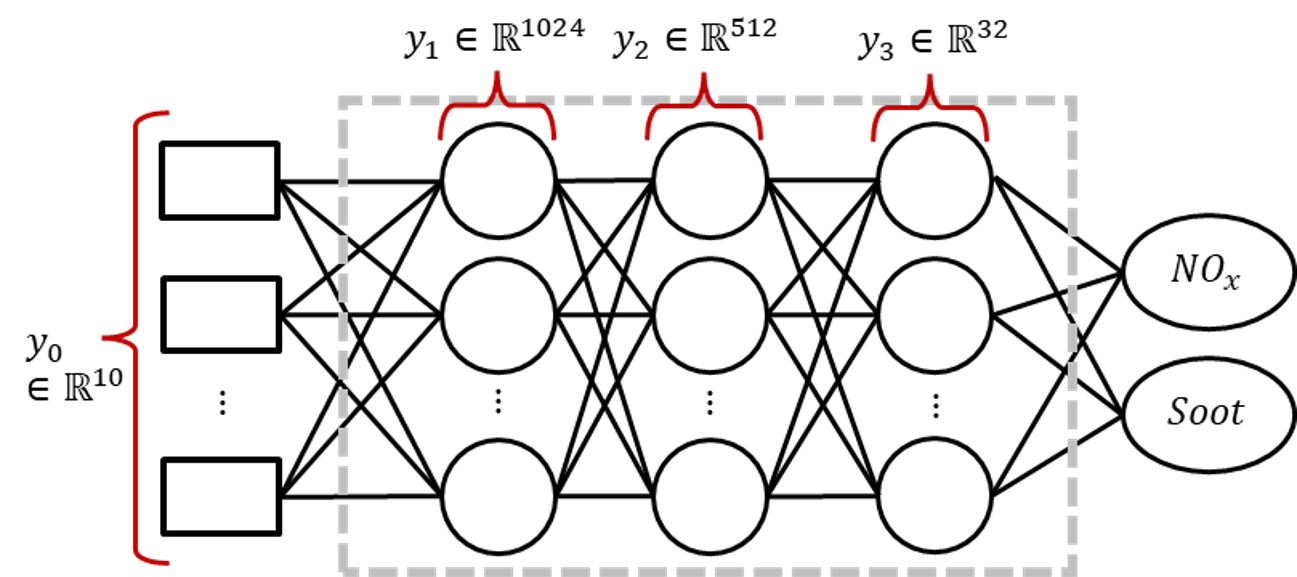}
\caption{Schematic of the developed multi-layer neural network diesel emissions model.}
\label{fig:nn_model}
\end{figure}

As in our previous work  \cite{IFAC2023}, we rely on a multi-layer FNN $f_{\text{FNN}}(\cdot)$ to represent engine $NOx$ and $Soot$ emissions.
The schematics of this FNN are shown in Figure~\ref{fig:nn_model}. 
The output of each layer is defined as
\begin{equation}\label{eq:nn_plant}
    y_i=\sigma_{ReLU}\left(W_iy_{i-1}+b_{i}\right),~~ i=1,\dots,4,
\end{equation}
where $W_i$ and $b_i$ are the network parameters of the $i${th} layer while $y_{0} \in\mathbb{R}^{10}$ and $y_{4}\in\mathbb{R}^{2}$ are the inputs and predicted emissions outputs, respectively.
Note that ReLU functions are used in (\ref{eq:nn_plant}) due to their good empirical performance in various neural network architectures, e.g., Residual Networks~\cite{resnet}, Graph Neural Networks~\cite{gcn}, and Transformers~\cite{attention}. 

The transient and steady-state dynamometer datasets encompass respectively, $12001$ and $306$ data points.  After selecting the inputs and removing outliers as detailed in \cite{IFAC2023},  a single data set for FNN training has been formed where $i$th data point has the form $(y_0^{(i)}, y_4^{(i)})$. Here the $y_0^{(i)}$ corresponds to the measurements of injection pressure, main injection timing, main injection fuel flow rate, engine torque output, engine speed, intake manifold pressure, exhaust manifold pressure, mass air flow, EGR position, and VGT position.  The $y_4^{(i)}$ corresponds to the measurements of the oxides of nitrogen ($NOx$) and $Soot$ emissions. 

\begin{figure}[ht!]
\centering
\includegraphics[width=0.48\textwidth]{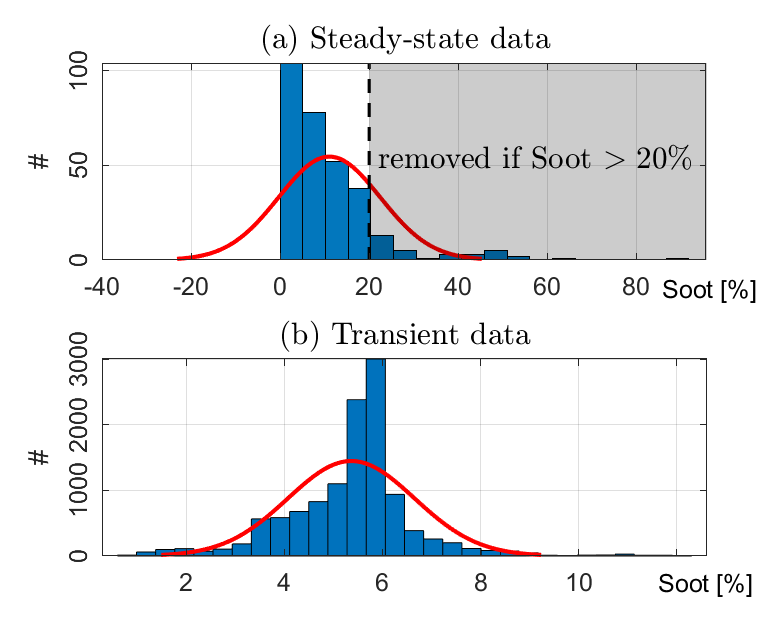}
\caption{Histogram of $Soot$ measurements with a distribution fit for (a) steady-state data and (b) transient data}
\label{fig:soot_histogram}
\end{figure}

The merging of transient and steady-state dynamometer datasets deserved additional consideration as the transient data set was collected with a nominal (non-MPC type) controller running and contained points close to those that would be encountered in closed-loop transient operation over the driving cycle with Soot values varying only over a small range which was achieved by fuel limiting.  On the other hand, the steady-state dataset has resulted from open-loop steady-state dynamometer mapping and contained actuator setting and Soot values over a much larger range (see Figure~\ref{fig:soot_histogram}).  At the same time, these points were from steady state engine operation in stabilized engine conditions and many of the points with larger Soot values would not be encountered in closed-loop operation once the controller has been calibrated. 

Using all points for FNN training would have been preferable, at first glance. However, in practice, it has resulted in an FNN model overemphasizing matching peaks in Soot present in the steady-state data set and over-predicting Soot peaks in the range expected during closed-loop operation. On the other hand, developing the model only from a transient dataset would limit model extrapolation ability and, later, the ability of the EMPC controller to control Soot through adjustment of airpath set-points rather than only fueling rate (and hence lead to worse drivability).  Consequently, a compromise approach was chosen where we removed steady-state data points with $Soot$ value larger than $20\%$ from the training data set.


With the data set encompassing both transient and steady-state data, $\mathcal{D}= \left\{(y_0^{(i)}, y_4^{(i)})\right\}_i$, we now train an FNN to 
represent a mapping from $y_0^{(i)}$ to $y_4^{(i)}$ that minimizes a loss function, i.e., we solve the following optimization problem,
\begin{equation}
\begin{array}{c}
    \min\limits_{\theta=\left\{W_j,b_j\right\}_{j=1,\dots,4}} \mathcal{L}(\mathcal{D}|\theta), \\
    \mathcal{L}(\mathcal{D}|\theta) = \frac{1}{\abs{\mathcal{D}}}\sum\limits_{i}\norm{f_{\text{FNN}}(y_0^{(i)}|\theta)-y_4^{(i)}}_2^2,
\end{array}
\end{equation}
where $\theta$ is the vector of FNN parameters (weights), $\abs{\mathcal{D}}$ is the number of data points in the dataset $\mathcal{D}$, and $\mathcal{L}(\mathcal{D}|\theta)$ is the empirical mean-squared error (MSE) between the actual outputs $y_4^{(i)}$ and the predictions $f_{\text{FNN}}(y_0^{(i)})$.


For FNN training we use the {\tt Pytorch} package\cite{pytorch}.
The hyperparameters used in neural network training are typically selected through trial and error. In this work, we have developed a more systematic procedure to search for the hyperparameter settings for the Batch Stochastic Gradient Descent (SGD) algorithm with momentum. The Batch SGD algorithm admits the following iterative form
\begin{equation}
\begin{array}{c}
    \theta_{k+1} = \theta_k - \lambda d_k,\\
    d_k = \nabla_{\theta_k} \mathcal{L}(\mathcal{D}_k|\theta_{k}) + \rho d_{k-1},\\
    d_0 =  \nabla_{\theta_0} \mathcal{L}(\mathcal{D}_0|\theta_0),\\
\end{array}
\end{equation}
where the learning rate $\lambda$ and momentum $\rho$ are the hyperparameters, and $k$ is the index of current iteration. Here $\mathcal{D}_k, k=0,1,2,\dots$ are equal-size (the size is named batch size) and disjoint batch datasets such that $\mathcal{D}=\cup_k\mathcal{D}_k$.

\begin{figure}[ht!]
\centering
\includegraphics[width=0.48\textwidth]{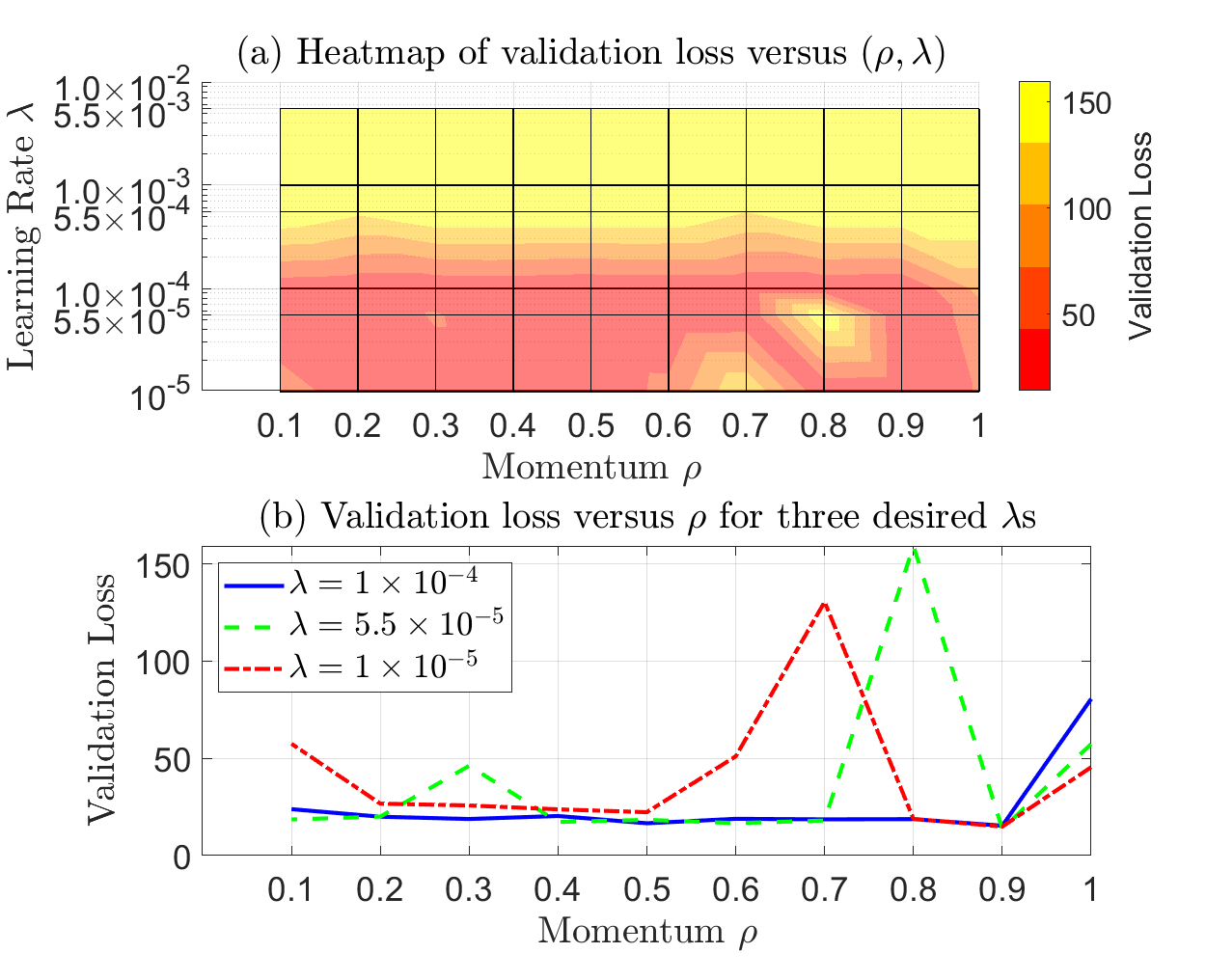}
\caption{Visualization of validation loss versus different combinations of momentum $\rho$ and learning rate $\lambda$: (a) Heatmap of validation loss versus $(\rho,\lambda)$. (b) Validation loss versus $\rho$ for three desired learning rates $\lambda$. }
\label{fig:hyperparameter_tuning}
\end{figure}

To select the hyperparameters $(\rho,\lambda)$, as illustrated in Figure~\ref{fig:hyperparameter_tuning}(a), we first define a region of interest in the 2D hyperparameter plane $\rho - \lambda$, and generate a coarse grid where each grid point is a plausible choice of $(\rho, \lambda)$ values for FNN training. For each grid point, we train an FNN using values of $(\rho, \lambda)$ at this grid for one epoch on the training dataset. Based on the validation dataset, we calculate the validation loss defined as the MSE between the actual and predicted emissions values using the trained FNN. 

We repeated the aforementioned procedure for all grid points, and the results are presented in Figure~\ref{fig:hyperparameter_tuning}. From Figure~\ref{fig:hyperparameter_tuning}(a), we can identify three learning rate values $\lambda=1\times10^{-4}, 5.5\times10^{-5}, 1\times10^{-5}$ that correspond to the smallest validation loss for all momentum values. After a detailed examination of Figure~\ref{fig:hyperparameter_tuning}(b) we observe that the choice of $\rho=0.9$ yields the lowest validation loss for all three desired learning rates. 
As a result, we choose $(\rho, \lambda)=(0.9, 1\times10^{-4})$ as our initial FNN training hyperparameter settings. We also set the learning rate decay to $\gamma=0.5$ for every $100$ epoch (i.e., we multiply the learning rate by $\gamma$ every $100$ epochs)  and we, finally, train the FNN for $1000$ epochs with the MSE loss function defined above and a batch size of $40$. Note that the chosen value of $\gamma$ is fairly large to enable exploration of FNN weight landscape during learning.

\begin{figure}[ht!]
\centering
\includegraphics[width=0.48\textwidth]{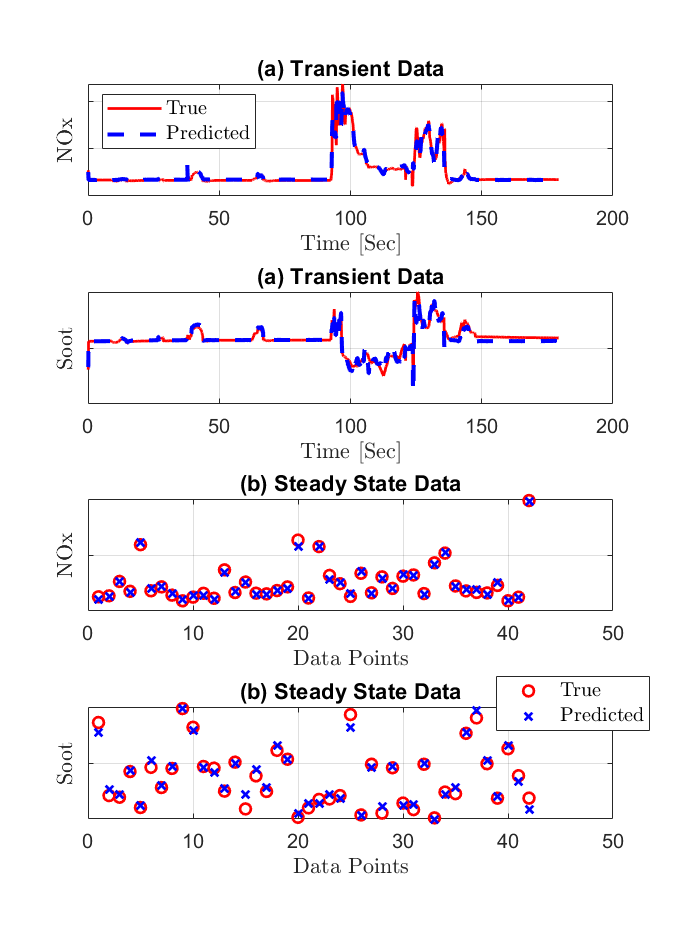}
\caption{The NN-based diesel emissions prediction performance of (a) transient and (b) steady-state data against the actual dynamometer data.}
\label{fig:nn_prediction}
\end{figure}

We compare the predicted emissions using our FNN with the actual dynamometer measurements in the testing part of the data set in Figure~\ref{fig:nn_prediction}. In Figure~\ref{fig:nn_prediction} and subsequent figures, we are not able to report the $y$-axis values in order to protect OEM proprietary data. For $NOx$ prediction, our FNN can achieve mean absolute prediction errors of $14.67\rm ppm$ for transient data and $25.39\rm ppm$ for steady-state data. For $Soot$ prediction, the mean absolute prediction errors are $0.21\%$ for transient data and $0.72\%$ for steady-state data, respectively. In Figure~\ref{fig:nn_prediction}, our predictions plotted in blue dash lines and crosses closely match the actual measurements (red lines and circles). This demonstrates good accuracy of our FNN model.

\subsection{Emissions Control-oriented Modeling}\label{subsec:RNN modeling}

To implement EMPC, a control-oriented model is needed for predicting the emissions response over the MPC optimization horizon. 
In our previous work \cite{IFAC2023}, we used a linear parameter-varying model (LPV) as a  control-oriented model. 
In this paper, a Recurrent neural network (RNN) model is developed and validated to improve the ability to predict the emissions. 
To address the larger computational footprint of the RNN, we adopted a simple RNN architecture with a small number of inputs, layers and neurons. 
In this model, $NOx$ and $Soot$ are selected as the model states ($x$). 
The current values of the  $NOx$ and $Soot$  emission outputs are assumed to be measured or accurately estimated in the ECU. Feedgas $NOx$ is often measured in diesel engines (even for the purpose of OBD)  while for $Soot$ an estimator (not necessarily the same as our control-oriented RNN) or a sensor based on advances in sensor technology is assumed (see e.g., \cite{malaczynski2017real}).
The input ($u$) for the RNN model encompasses the intake manifold pressure ($p_{\tt im}$), EGR rate ($\chi_{\tt egr}$), engine speed ($N_e$) and fuel injection rate ($w_{\tt inj}$). To that end, the prediction model has the following form, 
\begin{equation}  \label{eq:RNN model}
x_{k+1} = f_{\text{RNN}}(x_k,u_k),
\end{equation}
where $k$ denotes the discrete time instant.

\begin{figure}[ht!]
\centering
\includegraphics[width=0.4\textwidth]{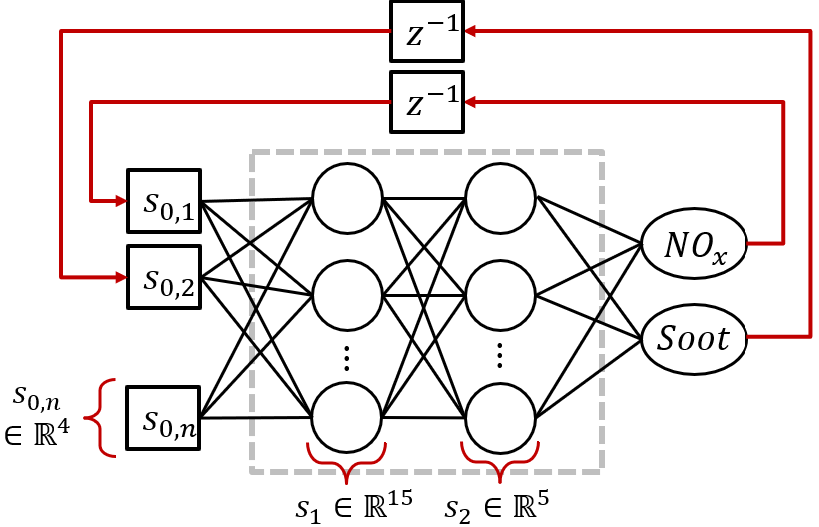}
\caption{Schematics of the multi-layer Recurrent Neural Network-based control-oriented model.}
\label{fig:RNN_model}
\end{figure}

Figure~\ref{fig:RNN_model} illustrates the adopted RNN model architecture with the two hidden layers of size $15$ and $5$ neurons, respectively. The output of each layer is defined by
\begin{equation}\label{eq:nn_plant2}
    s_i=\sigma_{Tanh}\left(\tilde{W}_is_{i-1}+\tilde{b}_{i}\right),~~ i=1,2,3,
\end{equation}
where $\tilde{W}_i$ and $\tilde{b}_i$ are the network parameters of the $i${th} layer while $[s_{0,1};s_{0,2}] \in\mathbb{R}^{2}$ are emission states (i.e., $x$), $s_{0,n} \in\mathbb{R}^{4}$ and $s_{3}\in\mathbb{R}^{2}$ are the inputs $u$ and the predicted emissions outputs, respectively.  Note that hyperbolic tangent functions $Tanh$ are used in (\ref{eq:nn_plant2}) rather than 
ReLU in (\ref{eq:nn_plant}) to ensure sufficient regularity of the model for first and second derivatives to be everywhere defined.

For identification of our RNN model, we use the data generated by a higher fidelity model consisting of the diesel engine airpath model in GT-Power 
and of FNN, which is used to predict $NOx$ and $Soot$ emissions based on GT-Power airpath model outputs.   The higher fidelity model is simulated over FTP and WHTC transient cycles to generate the data for RNN identification.   

\begin{figure}[ht!]
\centering
\includegraphics[width=0.48\textwidth]{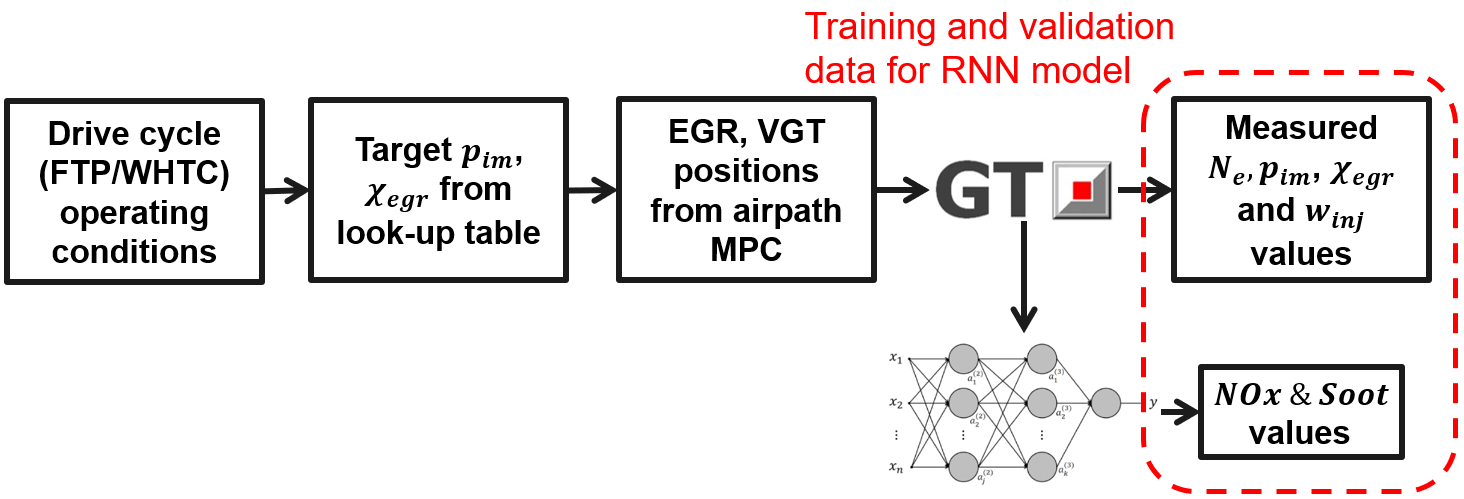}
\caption{Data generation procedure for identification of control-oriented RNN model of diesel emissions.  Here $N_e$ is the engine speed, $p_{\tt im}$ is the intake manifold pressure, $\chi_{\tt egr}$ is the EGR rate, and $w_{\tt inj}$ is the fuel injection quantity.}
\label{fig:RNN SID}
\end{figure}

The data generation procedure is summarized in Figure~\ref{fig:RNN SID}. The EGR valve and VGT positions are computed from a look-up table as functions of engine speed and fueling rate.  
Note that the airpath controller is not used as a part of this simulation.  Such an approach is beneficial in that it uncouples the identification of RNN for EMPC from the need to have a fully calibrated airpath controller, thereby facilitating a modularized controller development process. 
The training of the RNN employs the minimization of Mean Square Error (MSE) loss across the entire prediction horizon for each time step. This prediction horizon is aligned with the one utilized in our subsequent control simulation. The hyperparameters were tuned using the same process in Section~\ref{subsec:fnn_modeling}. The  data points used for training, validation, and testing have been chosen in proportion to the ratio of 70\%:15\%:15\%.

\begin{figure}[h!]
\centering
\includegraphics[width=0.48\textwidth]{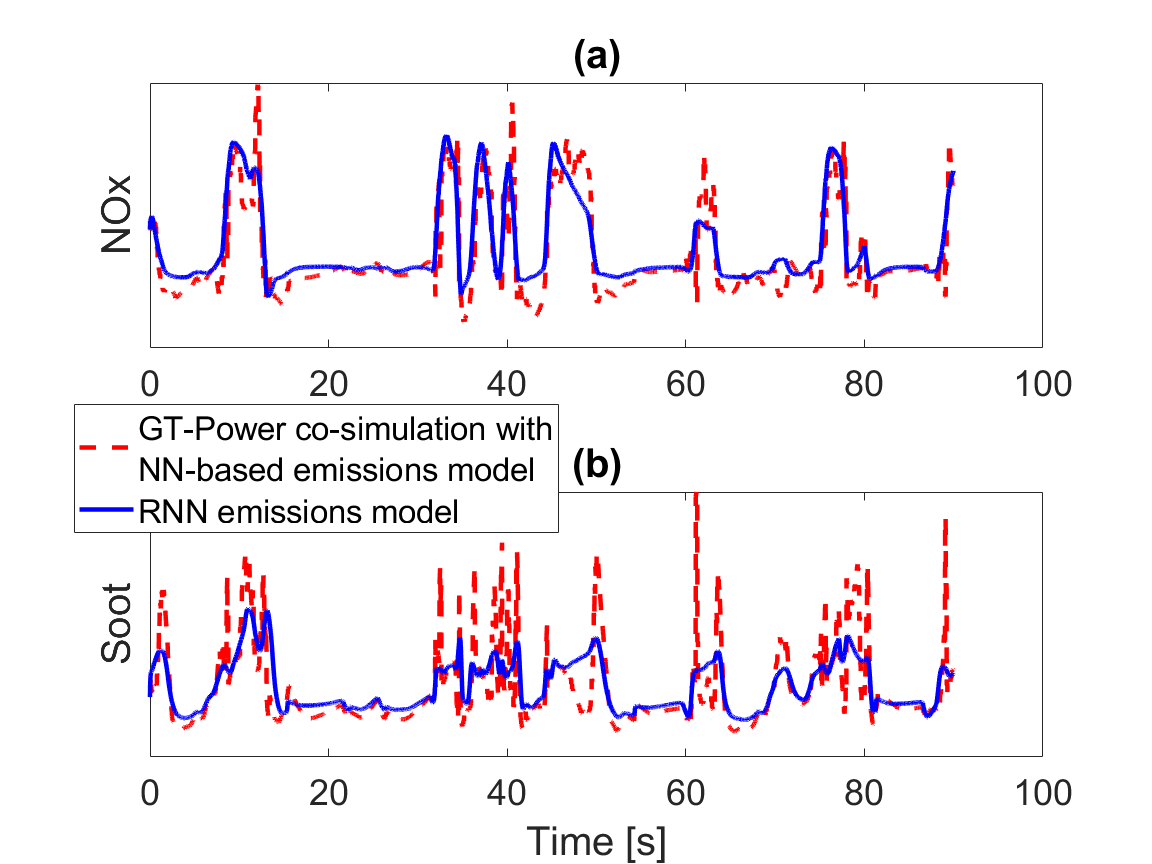}
\caption{The comparison between control-oriented RNN emissions model co-simulated with the GT-Power model and higher fidelity FNN emissions model over the last 15\% of WHTC driving cycle: (a) $NOx$, and (b) $Soot$.}
\label{fig:LPV validation}
\end{figure}

Figure~\ref{fig:LPV validation} compares the emissions predicted by the RNN model versus the FNN model.  These simulations are based on the last 15\% data of the WHTC driving cycle, which has not been used for RNN training. Both emission models are co-simulated with the GT-Power airpath model and respond to the same input trajectories of engine speed, fuel injection rate, target intake manifold pressure, and target EGR rate. 
The trajectories of the target intake manifold pressure and target EGR rate for these simulations were generated from
look-up tables that specify desired $p_{\tt im}$
and $\chi_{\tt egr}$ in steady-state as functions of the engine speed and fuel injection rate.

The average errors in $NOx$ and $Soot$ prediction for the RNN model as compared to the higher-fidelity FNN model are less than $43.3~ppm$ and 1.21\%, respectively. Larger errors are observed in $Soot$ prediction 
Although these transient errors could be reduced using higher complexity neural network models, as we will illustrate in the following sections, the proposed relatively simple RNN model effectively supports EMPC design.

\section{Economic MPC Design for Control of Engine Emissions}\label{sec:controller design}

\begin{figure}[ht!]
\centering
\includegraphics[width=0.48\textwidth]{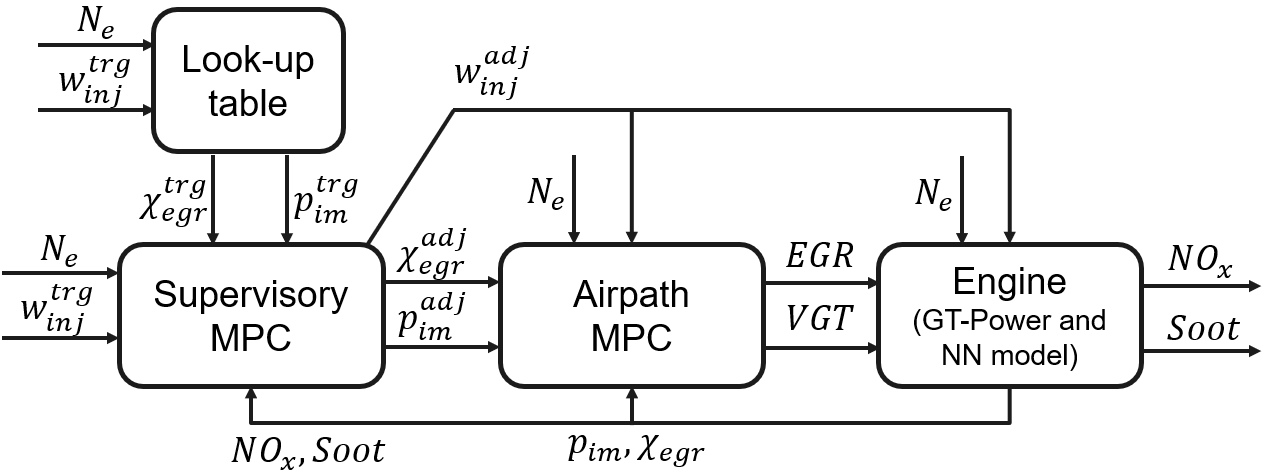}
\caption{The block diagram of the integrated emissions and airpath control system. 
}
\label{fig:structure}
\end{figure}

The overall schematics of the control system considered in this paper are shown in Figure~\ref{fig:structure}. 
Given engine speed ($N_{\tt e}$) and fuel injection rate target ($w_{\tt inj}^{trg}$), the outer-loop EMPC controller generates adjusted target values for intake manifold pressure ($p_{\tt im}^{adj}$), EGR rate ($\chi_{\tt egr}^{adj}$) and fuel injection command ($w_{\tt inj}^{adj}$) to both reduce $NOx$ emissions and enforce the $Soot$ constraint. While the fuel injection rate directly affects the engine operation, the adjusted intake manifold pressure and EGR rate target values are passed to the airpath MPC. The  airpath MPC  then coordinates the EGR valve and VGT positions to track these adjusted targets.

The design and tuning of the airpath MPC have been detailed in \cite{ZHANG2022181}.
Our airpath MPC includes a feedforward (FF) MPC, to provide fast transient response, and a feedback (FB) rate-based MPC
that provides integral action and ensures offset-free tracking.

In EMPC design, we use the model (\ref{eq:RNN model}) as the prediction model but with the input set to $u_k=[p_{\tt im}^{ adj},~\chi_{\tt egr}^{adj},~w_{\tt inj}^{adj}]^{\sf T}$, i.e., with intake manifold pressure, EGR rate and fueling rate replaced by the airpath MPC set-points.

Rate-based MPC is used to ensure offset-free tracking and to compensate for model mismatch. Its design and properties are well-understood in the linear quadratic MPC case, see e.g., \cite{wang2004tutorial,huang2016rate}. 
For nonlinear MPC of diesel engines, rate-based design  has been considered in~\cite{huang2016low} where, under certain assumptions, it was shown that the closed-loop system resulting from the rate-based formulation maintains the desired equilibrium despite offsets. 
Our RNN-based EMPC for emissions control also adopts a rate-based formulation.


First, in reference to the model (\ref{eq:RNN model}), we define rate  variables $\Delta x_{k}$ and $\Delta u_{k}$ as $$\Delta x_{k} = x_{k} - x_{k-1}, \quad \Delta u_{k} = u_{k} - u_{k-1},$$
that correspond to the increments in the state and control variables, respectively. 
Then, according to the prediction model (\ref{eq:RNN model}),
\begin{equation} \label{eq:rate-based RNN model}
\Delta x_{k+1} = x_{k+1}-x_{k} = f_{\text{RNN}}(x_{k},u_{k}) - x_{k}.
\end{equation}
We let $x_{j|k}$ and $u_{j|k}$ denote the predicted state and control values at time step $j,$
$0 \leq j \leq N$, over the prediction horizon when the prediction is made at the time step $k$. Then,
to be able to impose constraints on $x_{j|k}$ and $u_{j|k}$ and compute the MPC cost that penalizes the rate of change of input, $\Delta u_{k}$, we define the augmented state vector,
$$x_{j|k}^{\tt ext} = \left[\begin{array}{cccc} \Delta x_{j|k}^{\sf T}, & x_{j-1|k}^{\sf T}, & u_{j-1|k}^{\sf T} \end{array}\right]^{\sf T}.$$
Using  \eqref{eq:rate-based RNN model} it follows that
\begin{align}
    &x_{j+1|k}^{\tt ext} = \begin{bmatrix}
    f_{\text{RNN}}(x_{j-1|k}+\Delta x_{j|k},u_{j-1|k}+\Delta u_{j|k}) -\\ \quad\quad\quad\quad\quad\quad\quad\quad\quad\quad (x_{j-1|k}+\Delta x_{j|k}) \\
    x_{j-1|k}+\Delta x_{j|k}\\
    u_{j-1|k}+\Delta u_{j|k}
    \end{bmatrix}\label{eq:extended_system},
\end{align}
where $\Delta u_{j|k}$ is the control input in the extended system
(\ref{eq:extended_system}).

The EMPC design takes into account safety and drivability, ensuring that the fuel injection rate adheres to specified upper and lower limits, while also addressing emission requirements such as $NOx$ reduction and  $Soot$ constraints. At each time step, the following optimal control problem is solved to optimize the sequence of set-points for the intake manifold pressure, EGR rate, and fuel injection rate:
\begin{subequations}\label{eq:modified emission MPC}
\begin{multline}
    \min_{\Delta u_{0|k},...,\Delta u_{N-1|k},\epsilon_k} J(\Delta u,\epsilon,\rho) = \sum_{j=0}^{N} l(\Delta u_{j|k},\epsilon_{j|k},\rho_k)
\tag{\ref{eq:modified emission MPC}}
\end{multline}
subject to
\begin{align} 
    &\Delta x_{j+1|k} = x_{j+1|k}-x_{j|k} = f(x_{j|k},u_{j|k}) - x_{j|k},\\
    &u_{j|k} = u_{j-1|k} + \Delta u_{j|k}, j = {0}, \ldots, N-1,\\ 
    &{Soot}_{j|k} \leq {Soot}^{lim} + \epsilon_{j|k}, j = 1, \ldots, N,\\
    &0.9w_{\tt inj}^{trg}\leq {w_{\tt inj}^{adj}}_{j|k} \leq w_{\tt inj}^{trg}, j = 0,\ldots, N-1,\\
    &\epsilon_{j|k} \geq 0, j = 0,\ldots, N-1,
\end{align}
\end{subequations}
where $N$ is the prediction horizon and $\epsilon_k$ is the slack variable introduced to avoid infeasibility of the $Soot$ constraints. The stage cost function $l$ is defined as
\begin{align*}
    l(\Delta u,\epsilon,\rho) = &\alpha(p_{\tt im}^{trg}(\rho)-p_{\tt im}^{adj})^2 + \beta(\chi_{\tt egr}^{trg}(\rho)-\chi_{\tt egr}^{adj})^2 + \\
    &\gamma(w_{\tt inj}^{trg}-w_{\tt inj}^{adj}) + \eta {NOx} + \zeta \epsilon + \Delta u^{\sf T}R\Delta u
\end{align*}
where $\alpha,\beta,\gamma,\eta,\zeta > 0$ and $R > 0$ are tuning parameters and reflect tracking objectives for the intake manifold pressure target, EGR rate target, and fuel injection rate, a penalty on $NOx$ value, a penalty to soften the $Soot$ constraint to guarantee feasibility and a damping term. The cost terms for fuel tracking, $NOx$, and slack variables use 1-norm penalties since they are more robust to ill-conditioning compared to quadratic penalties \cite{liao2020model}, facilitating the solution of \eqref{eq:modified emission MPC} numerically. 

Once the solution to \eqref{eq:modified emission MPC} is determined,
the control is informed by the first move  $\Delta u_{0|k}^* $ 
of the optimal control increment sequence, i.e.,
$$u_{k} = u_{k-1} + \Delta u_{0|k}^*. $$

\section{Simulation Results and Discussions}\label{sec:simulations} 

The EMPC implementation employs the RNN model established in Section \ref{subsec:RNN modeling}, and incorporates both soft state constraints and hard control constraints. For MPC implementation and prototyping the package {\tt MPCTools} \cite{risbeck2016mpctools} is employed. The IPOPT solver is utilized for the numerical solution of the optimization problem in EMPC. Notably, {\tt MPCTools} offers the ability to generate symbolic expressions of the RNN model, facilitating the computation of Hessians and derivatives essential for the optimization problem.

We examine four different tuning scenarios to explore how emissions can be modified based on varying control objectives:

\begin{itemize}
  \item \textbf{EMPC-A}: Low $NOx$ penalty $\eta$ without $Soot$ limit;
  \item \textbf{EMPC-B}: High $NOx$ penalty $\eta$ without $Soot$ limit;
  \item \textbf{EMPC-C}: Low $NOx$ penalty $\eta$ with $Soot$ limit;
  \item \textbf{EMPC-D}: High $NOx$ penalty $\eta$ with $Soot$ limit.
\end{itemize}

A baseline scenario without EMPC (\textbf{w/o EMPC}) was also defined, in which the adjusted targets for 
$p_{\tt im}$, $\chi_{\tt egr}$ are the same as the values from look-up table while $w_{\tt inj}$ is the same as defined by the operating condition.

\subsection{Case Study with Steps/Ramps}

\begin{figure}[h!]
\centering
\includegraphics[width=0.45\textwidth]{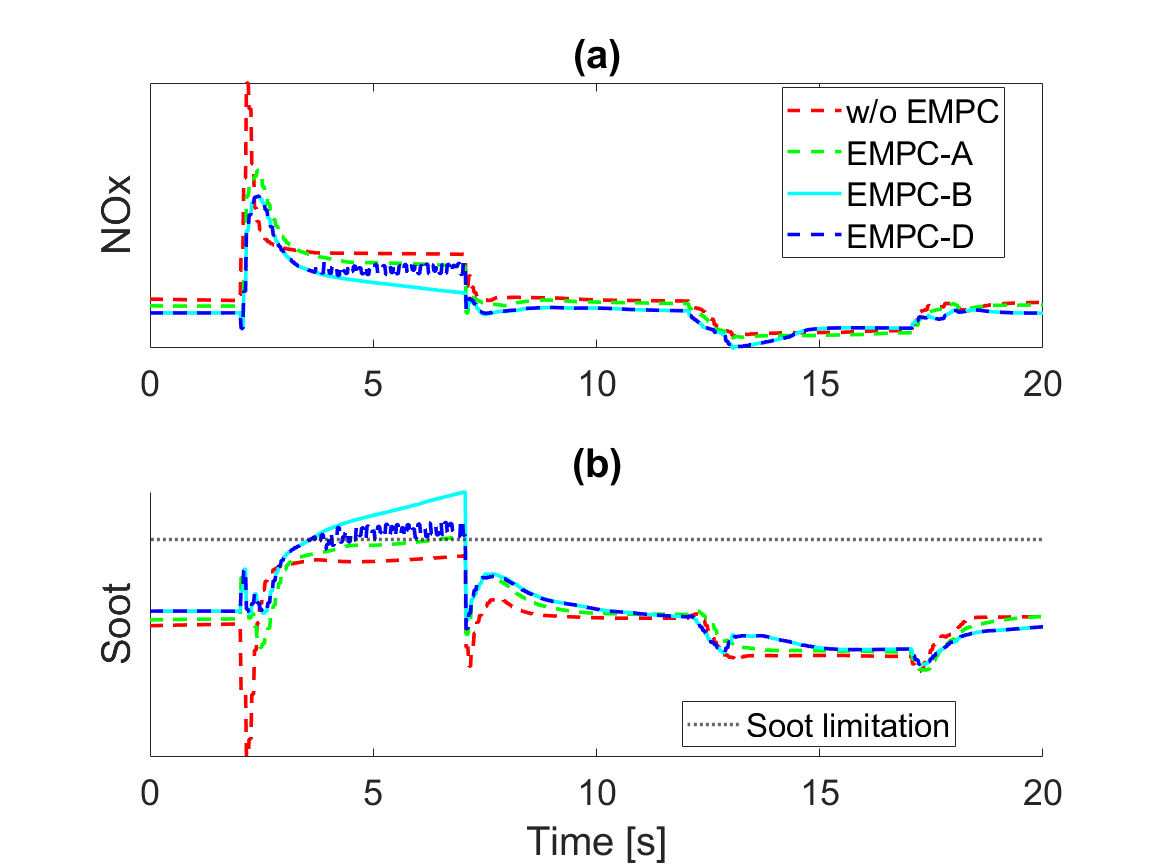}
\caption{Comparison of (a) $NOx$ and (b) $Soot$ with different $NOx$ penalties during {first, a fuel ($w_{\tt inj}$) tip-in and tip-out and then, a ramp change of $N_{\tt e}$.} 
}
\label{fig:case_study}
\end{figure}

\begin{figure}[h!]
\centering
\includegraphics[width=0.45\textwidth]{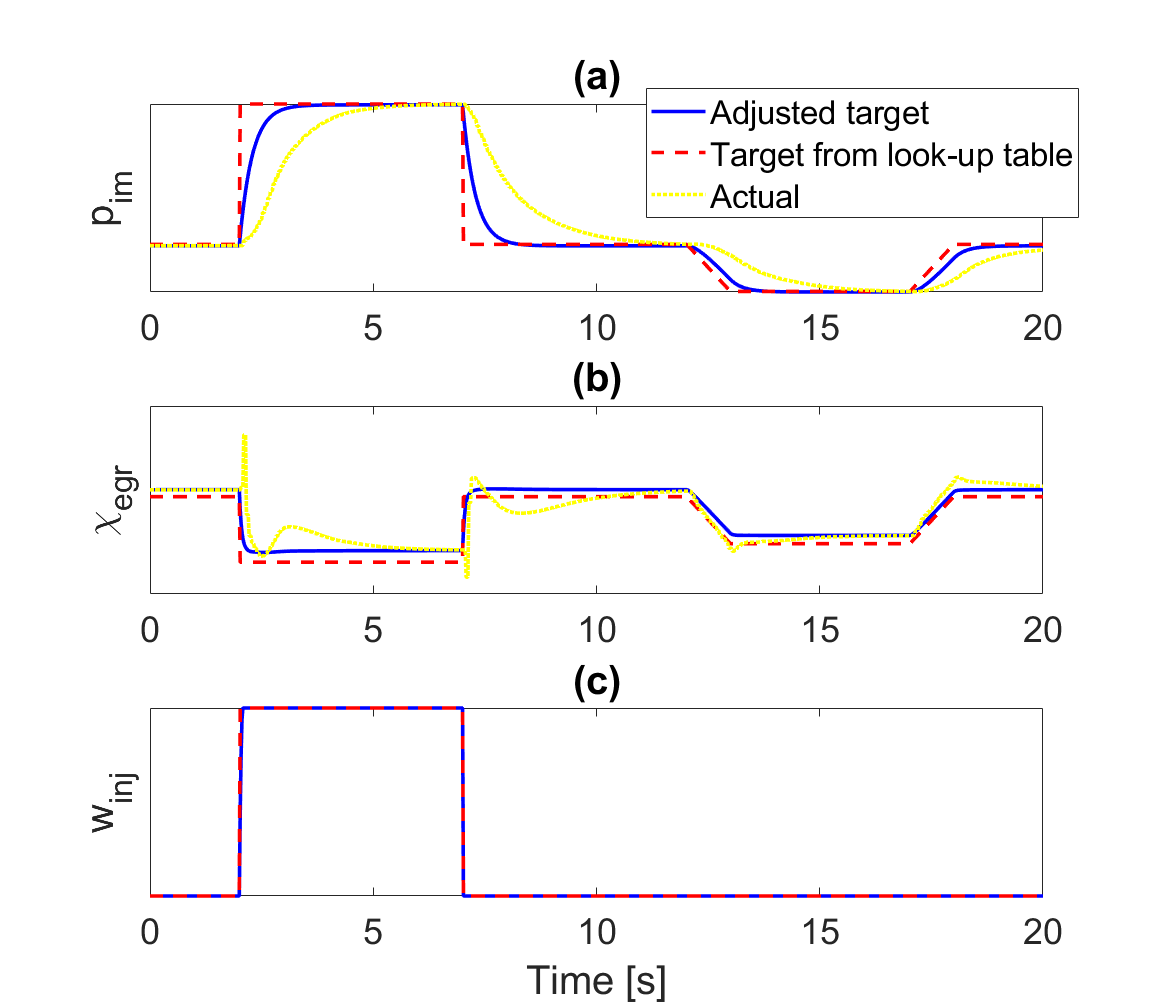}
\caption{Comparison of actual values, target values from look-up table, and adjusted target values by EMPC of the following physical quantities: (a) intake manifold pressure, (b) EGR rate and (c) fuel injection rate with \textbf{EMPC-B}.}
\label{fig:EMPC-B input}
\end{figure}

\begin{figure}[h!]
\centering
\includegraphics[width=0.45\textwidth]{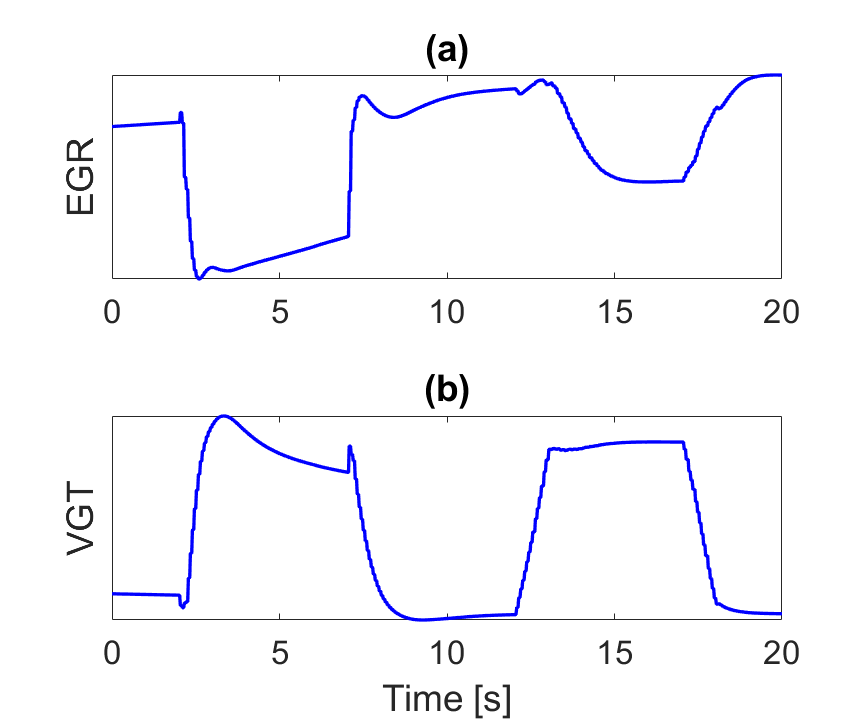}
\caption{(a) EGR and (b) VGT actuator behavior with \textbf{EMPC-B}.}
\label{fig:EMPC-B actuator}
\end{figure}

Figure~\ref{fig:case_study} presents simulation results applying 
 EMPC-A, EMPC-B, EMPC-D, and w/o EMPC controllers for a trace that encompasses a fuel tip-in, a fuel tip-out, and a subsequent ramp change in engine speed at a particular operating point. 
Given that the $Soot$ values for EMPC-A are mostly below the soft constraint, 
EMPC-C to EMPC-A exhibits a similar behavior and is not included in the subsequent figure and plot.
As detailed in Table~\ref{tbl:case study NOx}, the EMPC strategy reduces both cumulative and peak $NOx$ values, where the cumulative $NOx$ is defined as
\begin{equation}\label{equ:cumNOx}
    \int{w_{\tt ext}\times {\tt ppm}_{NOx}}{\tt dt}.
\end{equation}
In (\ref{equ:cumNOx}), $w_{\tt ext}$ is the exhaust flow and  ${\tt ppm}_{NOx}$ is the instantaneous $NOx$ concentration value in ppm. Figure~\ref{fig:EMPC-B input} illustrates the  time histories of the intake manifold pressure, EGR rate, and fuel injection rate and of the corresponding targets.  As these simulation results show, EMPC dynamically adjusts the targets for $p_{\tt im}$ and $\chi_{\tt egr}$ to facilitate $NOx$ reduction,
while the airpath MPC controls the EGR (\% open) and VGT (\% close) actuators as shown in Figure~\ref{fig:EMPC-B actuator} to accurately track these adjusted targets without steady-state errors.

Furthermore, as evident from Figure~\ref{fig:case_study}, upon introducing the $Soot$ constraint, EMPC-D is able to maintain $Soot$  close to or below a predetermined limit. Table~\ref{tbl:case study Soot} indicates that $NOx$ reduction may come at the expense of $Soot$ increase. However, the introduction of a $Soot$ constraint yields reductions in both average and peak $Soot$ values. Computing the ratio of the time during which $Soot$ exceeds the specified limit over the total simulation time also confirms that imposing $Soot$ constraint results in $Soot$ reduction. 
Note that $Soot$ values for EMPC-D are lower versus EMPC-B, which illustrates the impact of imposing the $Soot$ limit.

\begin{figure}[h!]
\centering
\includegraphics[width=0.45\textwidth]{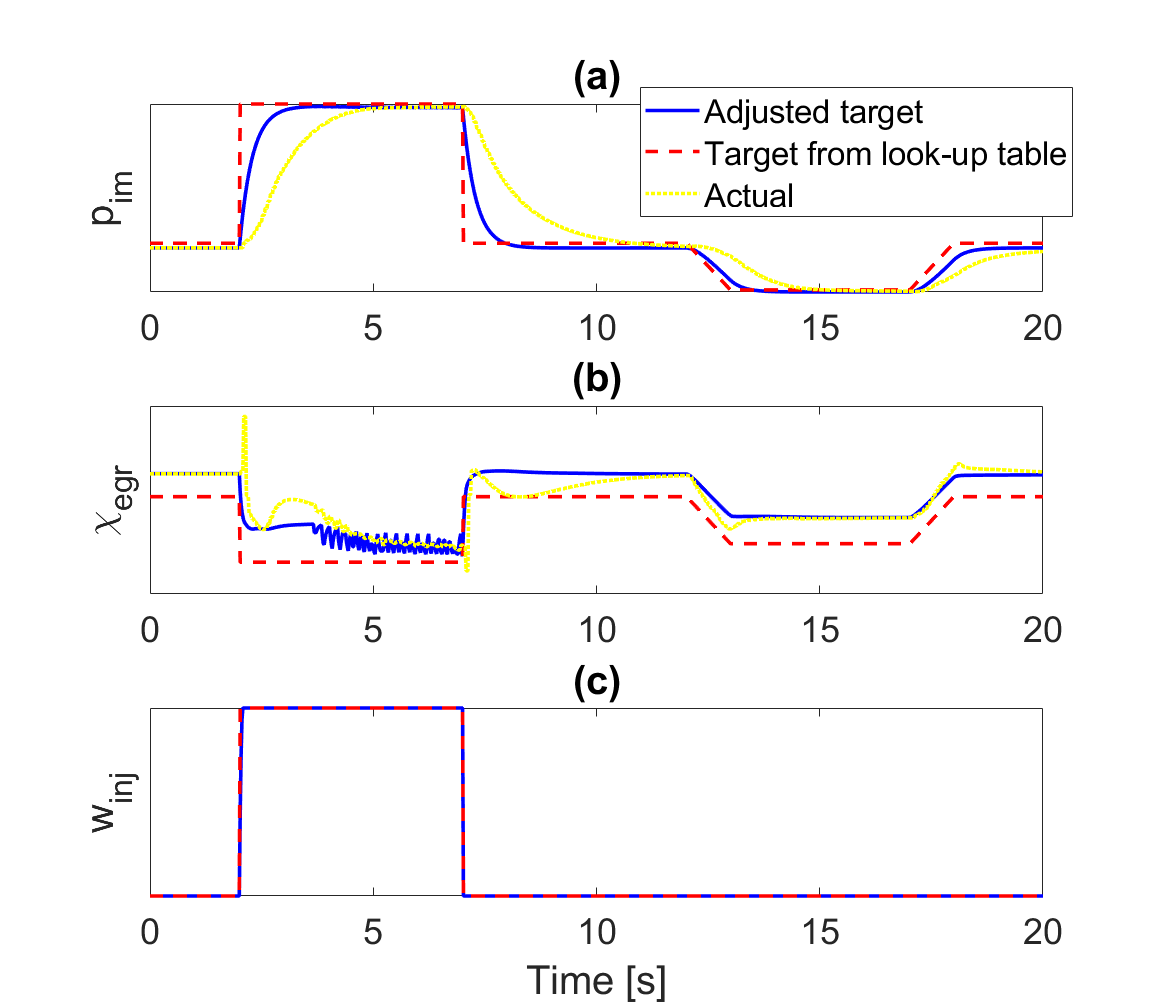}
\caption{Comparison of actual values, target values from look-up table, and adjusted target values by EMPC of the following physical quantities: (a) intake manifold pressure, (b) EGR rate and (c) fuel injection rate with \textbf{EMPC-D}.}
\label{fig:EMPC-D input}
\end{figure}

Oscillations in the adjusted targets $p_{\tt im}^{adj}$ and $\chi_{\tt 
egr}^{adj}$ are observed in Figure~\ref{fig:EMPC-D input} for EMPC-D when the $Soot$ output ``rides'' the constraint boundary.  These oscillations are attributed to the mismatch between the RNN model and the higher fidelity plant model. 
While these oscillations are undesirable as they can potentially increase the noise and wear, we note that the amplitude of these oscillations is small, and could be further reduced by increasing the damping term ($R$) in the MPC cost function.


\begin{table}[h!]
\caption{Comparison of the $NOx$ results for different EMPC scenarios.}
\label{tbl:case study NOx}
{\scriptsize
\begin{center}
\begin{tabular}{lll}
\toprule  
\makecell[l]{\textbf{MPC}} & \makecell[l]{\textbf{Cumulative $NOx$ }} & \makecell[l]{\textbf{Peak $NOx$}}\\
\midrule  
\textbf{w/o EMPC} & reference & reference\\ 
(baseline) &  & \\ \hline
\textbf{EMPC-A} & $\downarrow$ 1.941\% & $\downarrow$ 14.629\%\\\hline
\textbf{EMPC-B} & $\downarrow$ 6.116\% & $\downarrow$ 18.953\%\\\hline
\textbf{EMPC-D} & $\downarrow$ 4.994\% & $\downarrow$ 18.953\%\\
\bottomrule 
\end{tabular}
\end{center}}
\end{table}

\begin{table}[h!]
\caption{Comparison of the $Soot$ results for different EMPC scenarios.}
\label{tbl:case study Soot}
{\scriptsize
\begin{center}
\begin{tabular}{llll}
\toprule  
\makecell[l]{\textbf{MPC}} & \makecell[l]{\textbf{Average $Soot$}} & \makecell[l]{\textbf{Peak $Soot$}} & \makecell[l]{\textbf{${\text R}_{Soot > Soot^{lim}}$}}\\
\midrule  
\textbf{w/o EMPC} & reference & reference & 0\%\\ 
(baseline) &  &  &\\ \hline
\textbf{EMPC-A} & $\uparrow$ 2.301\% & $\uparrow$ 5.095\% & 2.997\%\\\hline
\textbf{EMPC-B} & $\uparrow$ 4.873\% & $\uparrow$ 16.482\% & 17.383\%\\\hline
\textbf{EMPC-D} & $\uparrow$ 3.740\% & $\uparrow$ 9.298\% & 14.386\%\\
\bottomrule 
\end{tabular}
\end{center}}
\end{table}

\subsection{Transient Drive Cycle Simulations}

\begin{figure}[h!]
\centering
\includegraphics[width=0.48\textwidth]{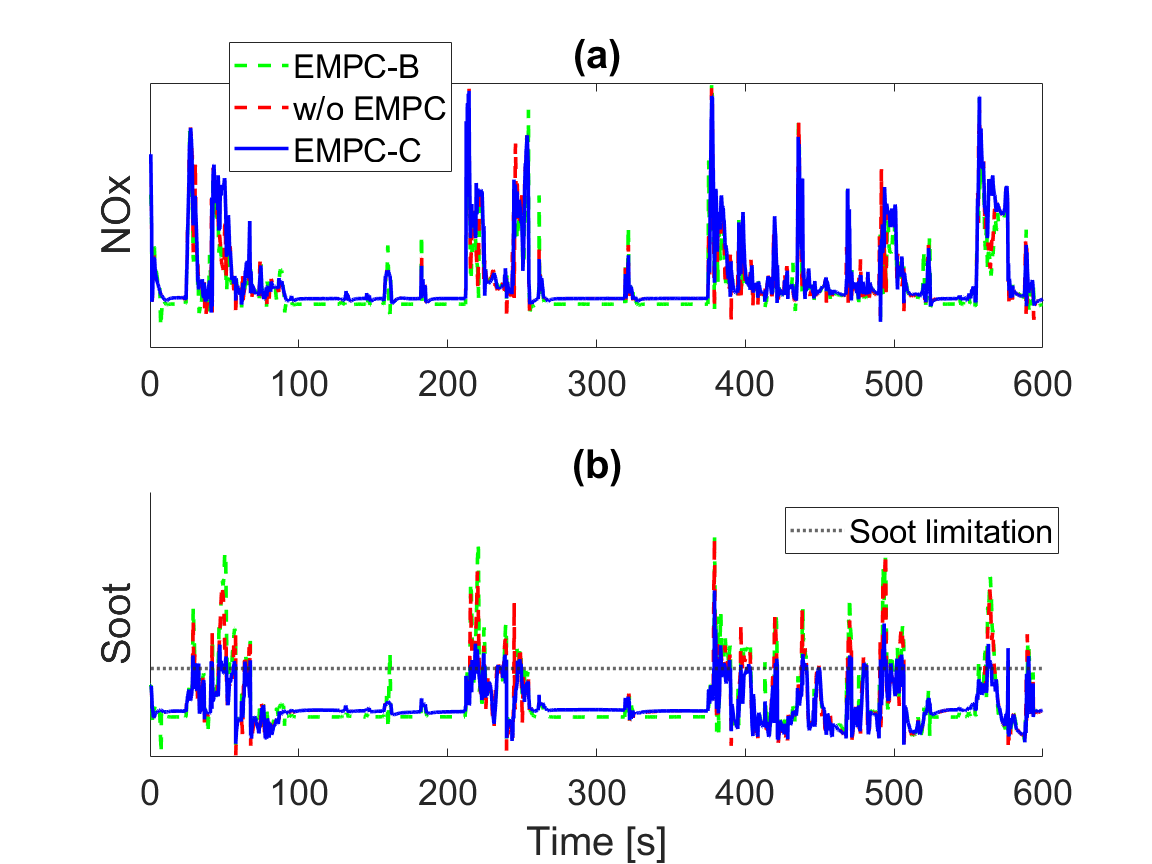}
\caption{Comparison of (a) $NOx$ and (b) $Soot$ control results over the FTP cycle.}
\label{fig:FTP compare}
\end{figure}

\begin{figure}[h!]
\centering
\includegraphics[width=0.48\textwidth]{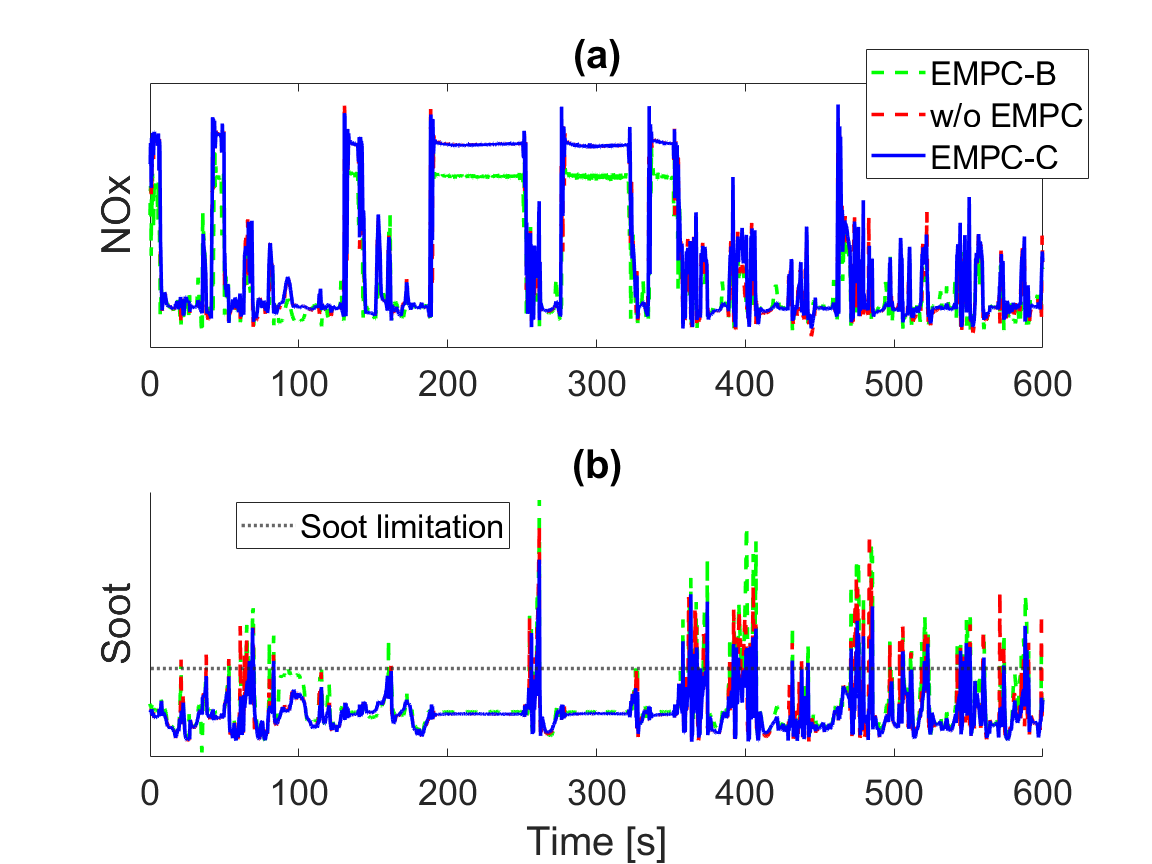}
\caption{Comparison of (a) $NOx$ and (b) $Soot$ control results over the WHTC cycle.}
\label{fig:WHTC compare}
\end{figure}

We next proceed to assess our integrated emissions and airpath control strategy in higher complexity simulation scenarios encompassing the Federal Test Procedure (FTP) and the World Harmonized Transient Cycle (WHTC). The results are presented in Tables~\ref{tbl:cycle NOx} and \ref{tbl:cycle Soot}. 
Figures~\ref{fig:FTP compare} and \ref{fig:WHTC compare} offer a  comparison of EMPC-B and EMPC-C against the baseline scenario over the FTP and WHTC cycles, respectively.

The application of EMPC-A leads to a slight reduction in $NOx$ emissions across both drive cycles. Subsequently, increasing the $NOx$ penalty, as seen for EMPC-B in Table~\ref{tbl:cycle NOx}, yields more reduction in cumulative $NOx$. However, this may lead to an increase in average and peak $Soot$ values across the drive cycles. 

Upon introducing a $Soot$ limit, EMPC-C 
reduces both average and peak $Soot$ values in both drive cycles compared to the baseline scenario. Furthermore, analysis of Figures \ref{fig:FTP compare}-(b) and \ref{fig:WHTC compare}-(b) demonstrates the substantial reduction of $Soot$ peaks exceeding its soft constraint, thereby showing EMPC-C's capability to diminish visible smoke emissions. The results corresponding to EMPC-D are exclusively presented in Tables~\ref{tbl:cycle NOx} and \ref{tbl:cycle Soot}, revealing a reduction in both cumulative $NOx$ and average $Soot$ concentrations when using a more aggressive EMPC strategy targeting both $NOx$ reduction and $Soot$ limits.

In addition, it's noteworthy that the total fuel consumption for all the cases in the FTP cycle and those in the WHTC cycle, when compared with the baseline, is reduced by 0.4\% and 0.6\%, respectively. This illustrates a more fuel-efficient control strategy without significantly compromising drivability.

\begin{table}[ht!]
\caption{Comparison of $NOx$ control results over FTP and WHTC cycles.}\vspace{-3pt}
\label{tbl:cycle NOx}
{\scriptsize
\begin{center}
\begin{tabular}{lll}
\toprule  
\makecell[l]{\textbf{MPC}} & \makecell[l]{\textbf{Cumulative $NOx$}} & \makecell[l]{\textbf{Peak $NOx$}}\\
\midrule  
\textbf{w/o EMPC} & reference & reference\\ 
(baseline) & FTP/WHTC & FTP/WHTC\\ \hline
 \textbf{EMPC-A} & $\downarrow$ 4.637\%/$\downarrow$ 1.913\% & $\downarrow$ 1.348\%/$\downarrow$ 5.514\%\\\hline
 \textbf{EMPC-B} & $\downarrow$ 9.463\%/$\downarrow$ 12.921\% & $\uparrow$ 0.064\%/$\downarrow$ 12.970\%\\\hline
 \textbf{EMPC-C} & $\uparrow$ 9.286\%/$\uparrow$ 4.962\% & $\downarrow$ 1.649\%/$\downarrow$ 0.719\%\\\hline
\textbf{EMPC-D} & $\downarrow$ 0.867\%/$\downarrow$ 8.299\% & $\downarrow$ 6.308\%/$\downarrow$ 12.968\%\\
\bottomrule 
\end{tabular}
\end{center}}
\end{table}

\begin{table}[ht!]
\caption{Comparison of $Soot$ control results over FTP and WHTC cycles.}\vspace{-3pt}
\label{tbl:cycle Soot}
{\scriptsize
\begin{center}
\begin{tabular}{llll}
\toprule  
\makecell[l]{\textbf{MPC}} & \makecell[l]{\textbf{Average $Soot$}} & \makecell[l]{\textbf{Peak $Soot$}} & \makecell[l]{\textbf{${\text R}_{Soot > Soot^{lim}}$}}\\
\midrule  
\textbf{w/o EMPC} & reference & reference & 10.21\%/5.22\%\\ 
(baseline) & FTP/WHTC & FTP/WHTC & FTP/WHTC\\ \hline
 \textbf{EMPC-A}  & $\downarrow$ 3.39\%/$\uparrow$ 3.44\% & $\downarrow$ 5.27\%/$\uparrow$ 12.62\% & 10.88\%/5.56\%\\\hline
 \textbf{EMPC-B}  & $\downarrow$ 0.77\%/$\uparrow$ 10.39\% & $\downarrow$ 0.41\%/$\uparrow$ 11.18\% & 11.67\%/6.81\%\\\hline
 \textbf{EMPC-C} & $\downarrow$ 4.56\%/$\downarrow$ 3.70\% & $\downarrow$ 24.69\%/$\downarrow$ 14.97\% & 6.47\%/2.93\%\\\hline
\textbf{EMPC-D}  & $\downarrow$ 7.425\%/$\uparrow$ 4.62\% & $\downarrow$ 16.07\%/$\downarrow$ 14.27\% & 7.46\%/3.94\%\\
\bottomrule 
\end{tabular}
\end{center}}
\end{table}

\section{Summary and Conclusions}\label{sec:conclusion}

The paper illustrated the use of neural networks both as part of simulation models and as a part of a prediction model for the implementation of an MPC-based diesel engine controller for emissions reduction.  The capability to shape engine feedgas emissions response has been demonstrated by adjusting weights in the cost function and constraints. To maintain low computational complexity we have adopted a relatively simple RNN architecture for the prediction model in EMPC. Further optimizing the computational characteristics of the resulting EMPC controller remains a topic for continuing research.


\bibliographystyle{IEEEtran}

\bibliography{IEEEabrv,reference}

\end{document}